# Comparison of measurement systems for assessing number- and mass-based particle filtration efficiency

Timothy A. Sipkens*, Joel C. Corbin, Triantafillos Koukoulas, Andrew Oldershaw, Thierry Lavoie, Jalal Norooz Oliaee, Fengshan Liu, Ian D. Leroux, Gregory J. Smallwood, Prem Lobo, Richard G. Green

*Metrology Research Centre, National Research Council Canada, Ottawa, ON, Canada*

## Abstract

The particle filtration efficiency (PFE) of a respirator or face mask is one of its key properties. While the physics of particle filtration results in the PFE being size-dependent, measurement standards are specified using a single, integrated PFE, for simplicity. This integrated PFE is commonly defined with respect to either the number (NBFE) or mass (MBFE) distribution of particles as a function of size. This relationship is non-trivial; it is influenced by both the shape of the particle distribution and the fact that multiple practical definitions of particle size are used. This manuscript discusses the relationship between NBFE and MBFE in detail, providing a guide to practitioners. Our discussion begins with a theoretical discussion of the underlying principles. We then present experimental results for a database of size-resolved PFE (SPFE) measurements for over 900 candidate respirators and filter media, including filter media with systematically varied properties and commercial samples that span the 20%−99.8% MBFE. The observed relationships between NBFE and MBFE are discussed in terms of the most-penetrating particle size (MPPS) and charge state of the media. For the NaCl particles used here, we observed that MBFEs were greater than NBFEs for charged materials and vice versa for uncharged materials. This relationship is observed because a shift from NBFE to MBFE weights the distribution towards larger sizes, while charged materials shift the MPPS to smaller sizes. Results are validated by comparing the output of a pair of TSI 8130As, a common instrument used in gauging standards compliance, to that of MBFEs computed from a system capable of measuring SPFE.

**Keywords**: particle filtration efficiency, respiratory protection, particle size, respiratory standards, PPE

## 1. Introduction

Respirators, face masks, and face coverings are critical to curbing respiratory disease transmission and providing personal protection against many forms of harmful aerosols, such as smoke, dust, and other pathogens. Their utility has been highlighted by the ongoing COVID-19 pandemic, where masks have significant utility as part of larger mitigation strategies (Hendrix 2020; Lyu and Wehby 2020; Stutt et al. 2020; Abboah-Offei et al. 2021; Bazant and Bush 2021; Gettings et al. 2021). The pandemic has also renewed interest in assessing the effectiveness of masks, including extensive consideration of common materials for use by the general public (Zangmeister et al. 2020; Zhao et al. 2020; Bagheri et al. 2021; Radney et al. 2021; Rogak et al. 2021; Zangmeister et al. 2021); the reuse of high performing masks (Lu et al. 2020; Ma et al. 2020; Rubio-Romero et al. 2020); the environmental impact of discarded masks (Fadare and Okoffo 2020; Dharmaraj et al. 2021; Hartanto and Mayasari 2021); and the testing methods and standards associated with various types of masks (Rengasamy et al. 2017; Corbin et al. 2021; LaRue et al. 2021; Zoller et al. 2021).

This paper is concerned with the material properties of candidate respirators or reference filtration media, chiefly the particle filtration efficiency (PFE), rather than consideration of fit. While fit is a very significant factor in mask performance (Rengasamy and Eimer 2011; Lei et al. 2013; Grima-Olmedo et al. 2014; Duncan et al. 2021), it is also challenging to measure effectively for the general populace. This is due, in part at least, to differences in the shape and features of individual faces and complex testing procedures (e.g., that require physical modifications to the mask, as in the OSHA test method (OSHA 2004), and a bivariate panel of volunteers). The accuracy of quantitative fit testing relies on how representative the test subjects are of the populace. Qualitative measures are often being used as a practical alternative. Further, the upper limit of effectiveness is governed by the material properties, requiring that standards continue to address material testing in addition to fit.

There are different types of PFE. The NIOSH TEB-APR-STP-0059 test method (NIOSH 2007), used for the certification of N95 respirators and hereafter referred to as the NIOSH N95 test method, traces back to the total amount of mass collected on a respirator. In doing so, this test method targets mass-based filtration efficiencies (MBFE). Thus, while commercial instruments available to gauge mask compliance (including the TSI 8130A, which features prominently in North America due to it being explicitly identified in the NIOSH test method, paragraph 3.1.1, and is used as a surrogate for commercial instruments in this work; ATI 100X; Gester GT-RA09; and Palas PMFT 1000) measure total light scattering from aerosol particles, these measurements are correlated to a MBFE via a calibration against gravimetric filter mass. These instruments can also be used in conjunction with other international standards that have similar requirements (Corbin et al. 2021). By contrast, the ASTM F2299 test method, which employs polystyrene latex (PSL) spheres, measures particle counts and thereby targets count- or number-based filtration efficiencies (NBFE). Academic/research studies often use systems composed of size-resolved particle counting instruments, such as scanning mobility particle sizers (SMPSs) (Li et al. 2012; Tang et al. 2018; Lu et al. 2020; Zangmeister et al. 2020; Corbin et al. 2021) and optical particle sizers (OPSs) (Tang et al. 2018; Rogak et al. 2021). The counting nature of these instruments is well suited to compute NBFE. However, the instruments also provide information on the size-resolved PFE (SPFE), which can be used to convert between NBFE and MBFE. The NBFE and MBFE will only be the same under certain conditions. Further, relating these two quantities is non-trivial, and the extent of the differences will depend on the size distribution, the shape of the SPFE curve, and the most penetrating particle size (MPPS). While noted as important (Zoller et al. 2021), this relationship has not seen much dedicated, experimental study, particularly across such a wide range of samples and into the sub-90% MBFE range.

Within this work, we investigate differences between these various quantities. We start by providing a detailed, theoretical discussion of particle size and the principles differentiating NBFE and MBFE. We then consider a research system – the Particle Filtration Efficiency Measurement System (PFEMS) (Smallwood et al. 2020; Corbin et al. 2021), composed of condensation particle counters and SMPSs – and report on measurements taken over more than a year, spanning a wide range of respirators and filtration media consisting of over 900 samples. We compute both NBFE and MBFE for this data and examine the relationship between the two quantities across the full range of samples. Of the materials tested, we also include a set of candidate reference filter media composed of melt-blown polypropylene where the charge state and basis weight are controlled. This allows for qualitative statements about the effect of the most common parameters that are adjusted in mask construction and their relation to the NBFE and MBFE. Overall, differences between NBFE and MBFE depend significantly on the charge state of the material. Finally, we compare the mass-based outputs from the size-resolved system to the output of two TSI 8130As, to assess the equivalence of these two characterization methods.

## 2. Abbreviations

*CMAD*. Count median aerodynamic diameter.

*CMD*. Count median mobility diameter (equal to GMD for number-based lognormal distributions). Also denoted by $d_g$.

*CPC*. Condensation particle counter.

*GMD*. Geometric mean diameter.

*GSD*. Geometric standard deviation.

*GSM*. Grams per square meter, in reference to basis weight.

*MBFE*. Mass-based PFE, integrated for a specific particle size distribution.

*MMAD*. Mass median aerodynamic diameter.

*MMD*. Mass median mobility diameter (equal to GMD for mass-based lognormal distributions).

*MPPS*. Most penetrating particle size.

*NBFE*. Number-based (or count-based) PFE, integrated for a specific particle size distribution.

*PFE*. Particle filtration efficiency, as a summary term for SPFE, NBFE, and MBFE.

*PFEMS*. The National Research Council Canada's Particle Filtration Efficiency Measurement System (Smallwood et al. 2020).

*PSL*. Polystyrene latex.

*SMPS*. Scanning mobility particle sizer.

*SPFE*. Size-resolved PFE, a set of PFE measurements as a function of particle size. Independent of particle size distribution used for evaluation.

*UQ*. Uncertainty quantification.

## 3. Background

A single aerosol distribution can be described by a range of particle characteristics (Hinds 1999b). The reported size and relevant measurement principle for particle filtration efficiency (PFE) data have varied in the literature, with inconsistent reporting in some cases. Aerosol particle size is a complex topic, for two reasons. First, particle transport in a gas is described by different equivalent sizes, depending on the context. Second, particles are often measured or described by reference to all particle sizes present in a given sample 'distribution'. These two points are expanded on in the following sections.

### 3.1. Particle sizes

Particle filtration is achieved through mechanical – including diffusion, impaction, and interception – and electrostatic mechanisms (Hinds 1999a; Tcharkhtchi et al. 2021). The mechanical mechanisms depend largely upon the particle size and face velocity, as well as fibre size (related to basis weight), whereas the electrostatic mechanism is almost solely dependent on the charge state of the filtration media (e.g., high performance melt-blown polypropylene materials typically achieve this high filtration and low pressure drop using electret materials).

In the context of diffusive capture by filters, particles behave according to their equivalent *mobility diameter*, $d_m$. The mobility diameter depends on the aerodynamic drag between the particle and the gas (DeCarlo et al. 2004). This is the quantity measured by the SMPSs within the PFEMS system used in this study (Corbin et al. 2021) and many others, e.g., (Bałazy et al. 2006; Lu et al. 2020; Zangmeister et al. 2020; Hao et al. 2021).

In the context of inertial motion or terminal velocity, as is relevant to the impaction and interception mechanisms, particles behave according to *aerodynamic diameter*. The aerodynamic diameter depends on the particle density and, inversely, on the aerodynamic drag. The precise relationship is given by DeCarlo et al. (2004). The aerodynamic diameter can be measured directly using inertial techniques, such as impactors. It can also be measured by accelerating particles to their terminal velocity and either measuring their velocity directly, as in laser-based aerodynamic particle sizers (TSI APS 3321) and various mass spectrometers (Pratt and Prather 2012) or by separating particles according to their time-of-flight, as in the Aerodynamic Aerosol Classifier (Issman et al. 2021; Payne et al. 2021) or Aerosizer (Qian et al. 1998). Larger, heavier particles are captured by filters and the human respiratory system according to their aerodynamic diameters (Kulkarni et al. 2011). We note that respiratory particles often have densities similar to water and thus have similar aerodynamic and mobility diameters. The same does not hold for sodium chloride, which is much denser than water, such that aerodynamic diameters for the same particles are often double or triple the mobility diameter.

A third diameter, the *optical diameter*, is relevant to measurements of particle filtration using photometers, such as that found in the TSI 8130A and optical particle counters (Rogak et al. 2021). The precise definition of this diameter is both material and device specific as it depends on the detection wavelength and collection angle.

Thus, particle sizes must be converted between mobility and aerodynamic size depending on the context. When a particle's properties are well known, a conversion between its mobility and aerodynamic size can be easily performed. Optical diameters can also easily be predicted if the instrument parameters are well known.

### 3.2. Particle distributions and distribution moments

A collection of particles may be described in terms of its total number, total mass, or other property (e.g., amount of light-scattering). More commonly, the same property may be divided into a *size distribution*, denoting the contribution of particle of different sizes in a population to the respective total. In aerosol science, distributions are typically denoted using a differential, with the quantity of particles (e.g., count, mass) in the numerator and a measure of the particle size in the denominator. For example, d$N$/dlog$d_m$ is used to denote the number distribution resolved with respect to mobility diameter. The distribution for the same population of particles changes depending on the chosen parameters. For example, in the case of numerator, the contribution of larger particles to the total number is much smaller than the contribution of those particles to the total mass. Consequently, mass distributions (e.g., d$M$/dlog$d_m$) typically peak at larger sizes than number-based distributions. Size distributions are often specified using a range of distribution moments, including count median mobility diameter (CMD) for number distributions and mass median mobility diameter (MMD) for mass distributions. The distribution moments themselves can also be expressed in terms of the different particle sizes (when the denominator of the distribution is changed). For instance, the CMD and MMD can also be expressed as a count median aerodynamic diameter (CMAD) and mass median aerodynamic diameter (MMAD), respectively.

In the context of this work, we note that the scaling of mass with diameter can typically be formulated using a power law with respect to diameter,

$$m = m_0 d^\zeta,\qquad(1)$$

where $\zeta$ is a power applied to the count-based diameter to achieve the desired output and $m_0$ is a pre-factor. For the mass of spheres and $d = d_m$, we note that $\zeta = 3$ and $m_0 = \pi\rho/6$, where $\rho$ is the particle's density. For Eq. (1) and lognormal distributions, the MMD is related to the CMD by the Hatch-Choate relation (Heintzenberg 1994; Hinds 1999b),

$$d_\zeta = d_g \exp\left[\zeta\left(\ln\sigma_g\right)^2\right].\qquad(2)$$

where $d_g$ is the CMD and $d_\zeta$ is the MMD. Note that the MMD does not depend on the pre-factor (i.e., is independent of the particle density) but does depend on the distribution width. For infinitesimally narrow distributions, $\sigma_g = 1$ (i.e., the target of testing with PSL particles), the CMD and MMD are identical, an indication that the number and mass distributions are also identical. Note that, provided the mass is related to the diameter via a power law, it can be shown that the number and mass distribution widths are identical, only differing in terms of the center of the distribution.

Figure 1 demonstrates the equivalences of these properties for sodium chloride with a CMD of $d_g = 75$ nm and $\sigma_g = 1.8$ (close to the upper limit of the NIOSH N95 test method). Changes to the denominator (e.g., changing from mobility to aerodynamic diameter) changes the values on the $x$-axis. Changes to the numerator shift the distribution (on any axis), a consequence of reweighting the importance of the particle contributions to a given total.

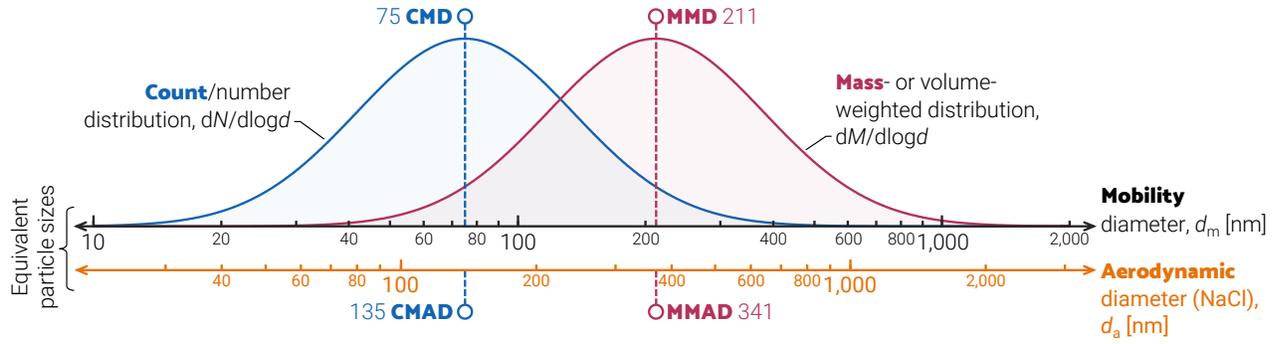

**Figure 1.** Schematic demonstrating the different particle sizes and types of size distributions for the same population of sodium chloride particles, assumed spherical with $d_g = 75$ nm, and $\sigma_g = 1.8$ (with respect to mobility diameter). Distributions are lognormal, such that they appear normal on the logarithmic scale here. The second horizontal axis shows the equivalent values for aerodynamic diameter (e.g., a mobility diameter particle of 75 nm has an aerodynamic diameter of 117 nm), which involves a transformation of the horizontal axis. Wider distributions will result in a larger MMAD (MMAD = 380 nm at $\sigma_g = 1.86$, closer to the upper limit in the NIOSH TEB-APR-STP-0059 test method).

### 3.3. Different types of filtration efficiency

Based on these concepts, we distinguish between three different types of particle filtration efficiency (PFE): size-resolved filtration efficiency (SPFE), number-based filtration efficiency (NBFE), and mass-based filtration efficiency (MBFE).

The *SPFE* requires a pair of size-resolved measurements, which is more common in academic studies. Then, the size-resolved penetration, $P_i$, and SPFE, $\eta_i$, at the *i*th particle size, $d_i$, are computed by

$$P_i = 1 - \eta_i = \frac{N_{(2,i)}}{N_{(1,i)}} \frac{1}{R_i},$$ (3)

where $N_{(1,i)}$ and $N_{(2,i)}$ are the number concentrations upstream and downstream of the respirator at standard temperature and pressure, respectively, and $R_i$ is a correction to account both for losses and for differences in instrument calibration between the upstream and downstream locations. The quantity $R_i$ is determined by making measurements without a respirator present. (This contribution will not be present for cases when a single downstream instrument measures both $N_{(1,i)}$ and $N_{(2,i)}$ using consecutive measurement with and without the sample present.)

The *NBFE*, by contrast, corresponds to the overall filtration for a *specific challenge aerosol*, averaged over the number distribution. This can be realized in one of two ways. First, one can integrate the size-resolved upstream and downstream number distributions, that is computing the total upstream particle concentration as

$$N_1 = \int \frac{dN_1}{d\log d} \cdot d\log d \approx \sum_i N_{1,i},$$ (4)

with an analogous expression downstream. Then, analogous to Eq. (3) but using total particle concentration measurements,

$$P_n = 1 - \eta_n = \frac{N_2}{N_1} \frac{1}{R}.$$ (5)

Alternatively, avoiding the losses and limited particle-size range associated with many size classifiers, one can measure the total number concentrations using an instrument that directly integrates the number distribution – for example using a pair of CPCs, as is available in the PFEMS – and compute the NBFE directly without the need to integrate as a post-processing step.

The *MBFE*, which is necessary for comparison against the TSI 8130A systems, is analogous to the NBFE but is averaged over the mass distribution instead of the number distribution, which amounts to the type of conversion described in Sec. 3.2. There are multiple ways to convert between NBFE and MBFE (Sipkens et al. Submitted). For instance, one can use numerical integration, e.g., Zoller et al. (2021). Here, MBFEs for the PFEMS are instead computed using a variant of Hatch-Choate analysis in conjunction with Eq. (1). In this case, the upstream and downstream distributions are assumed to be lognormal, and the overall mass concentration of particulate is given by,

$$M_1 = N_1 H_1 = N_1 \left\{ m_0 d_{g,1}^q \exp\left[ \frac{\zeta^2}{2} \left( \ln \sigma_{g,1} \right)^2 \right] \right\}, \tag{6}$$

where $H_1$ is the Hatch-Choate factor for the upstream size distribution; $d_{g,1}$ and $\sigma_{g,1}$ are the upstream geometric mean diameter (GMD), which is identical to the median for a lognormal distribution, and geometric standard deviation (GSD). (Relative to Eq. (2), this defines an integrated quantity instead of a distribution moment.) Assuming spheres, $\zeta = 3$, $m_0 = \pi\rho/6$, and

$$M_1 = N_1 \frac{\rho\pi}{6} d_{g,1}^3 \exp\left[ \frac{9}{2} \left( \ln \sigma_{g,1} \right)^2 \right], \tag{7}$$

where $\rho$ is taken here as the density of sodium chloride (2,160 kg/m³). An analogous conversion applies downstream. Then,

$$P_m = 1 - \eta_m = \frac{M_2}{M_1} \frac{1}{R} = \frac{H_2}{H_1} P_n = \frac{d_{g,2}^3}{d_{g,1}^3} \exp\left( \frac{9}{2} \left[ \left( \ln \sigma_{g,2} \right)^2 - \left( \ln \sigma_{g,1} \right)^2 \right] \right) P_n. \tag{8}$$

Note that, since $m_0$ contributes identically to $M_1$ and $M_2$, the mass-based penetration and filtration efficiency are independent of $m_0$ (and thus, for spheres, are independent of the particle density). Here, the GMD and GSD required for the Hatch-Choate factors are computed from the upstream and downstream number distributions via their statistical definitions (Hinds 1999b). This analysis is expected to be valid when the input particle size distribution and size-resolved penetration curve are approximately lognormal. One can also correct for $\zeta \neq 3$ in post-processing, provided that the distribution widths are available (see the Supplemental Information).

Overall, converting from NBFE to MBFE effectively changes the reported PFE from a measurement at 75 nm mobility diameter to a measurement around 220 to 380 nm MMD (or 357 to 593 nm MMAD, for $\sigma_g$ = 1.8 to 1.86). This occurs because the distribution is reweighted towards larger particle sizes.

### 3.4. Theoretical results for the NBFE to MBFE conversion

The preceding statements allow for some theoretical discussion of the conversion from NBFE to MBFE (discussed only briefly elsewhere, e.g., (Qian et al. 1998)), which is demonstrated schematically in Figure 2. Universally, mass-based distributions are centered about a larger size than number-based distributions. Now, also consider penetration curves centered about a most penetrating particle size (MPPS). When the MPPS is smaller than the CMD, more particles pass through the filtration media around the center of the number-based distribution, which translates to MBFE > NBFE. Conversely, if the MPPS is larger than the MMD, more large particles pass through the filtration media, coinciding better with the center of the mass-based distribution, thus NBFE > MBFE. For the intermediate range, where the MPPS is between the CMD and MMD, there is a transition between these two conditions, with NBFE = MBFE when the MPPS is around the midpoint between CMD and MMD.

The magnitude of the difference between the NBFE and MBFE depends on quantities other than MPPS, such as the size-dependence of the penetration curve and the challenge aerosol distribution width. Generally, strongly-size dependent filtration media (size dependent penetrations) will result in larger differences between the NBFE and MBFE (Figure 2*b*, *c*), while media with flat (size-independent) penetrations across the distribution width will yield nearly identical NBFE and MBFE (Figure 2*a*). Similarly, if the particle size distributions are narrow, number and mass distributions become nearly coincident, requiring a very strong size dependent penetration to result in a difference between the NBFE and MBFE (culminating in NBFE and MBFE necessarily being the same for monodisperse particles).

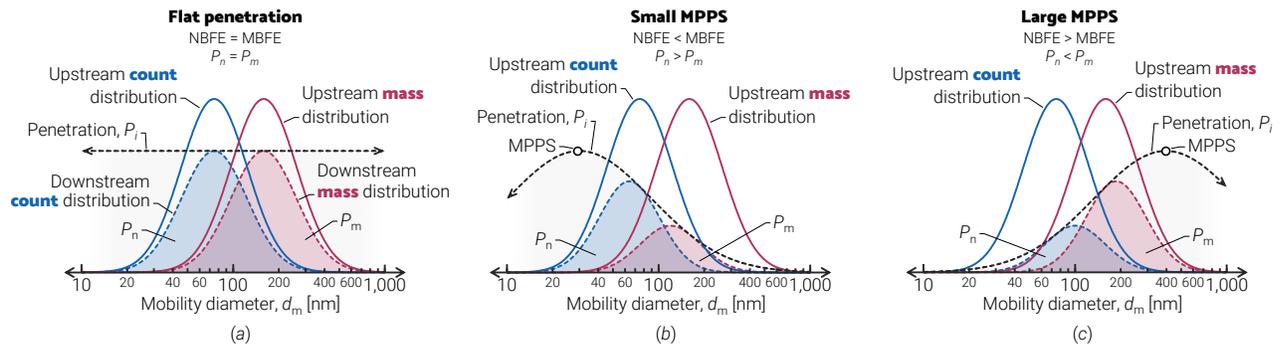

**Figure 2.** Three scenarios comparing NBFE and MBFE relative to the underlying size-resolved penetration, $P_i$, curves. The number- ($P_n$) and mass-based ($P_m$) penetrations result from integrating the product of the corresponding distribution and the size-resolved penetration curve, corresponding to the blue and red shaded regions, respectively. A size-independent penetration (*a*) results in NBFE = MBFE (the blue and red shaded regions are identical), while size-resolved $P_i$ (*b-c*) can yield differences between NBFE and MBFE.

# 4. Experimental details

## 4.1. PFEMS (number-based, size-resolved)

Figure 3 shows a schematic of the number-based PFEMS (Smallwood et al. 2020) at the National Research Council Canada. Due to the range of dates associated with this data (extending over a year and a half), some instruments were exchanged over the duration of these measurements, depending on availability. Despite this, the overall function remained constant, with comparison to a reference respirator showing consistency across the changes. The PFEMS contains an upstream nebulizer (TSI Collison 3076) used to generate a sodium chloride aerosol approximating that specified for the NIOSH TEB-APR-STP-0059 test method, typically with a geometric count mean diameter (CMD) of $d_{g,1} \sim 75$ nm and a geometric standard deviation (GSD) that varied between $\sigma_{g,1} \sim 1.65$ and $1.85$ (earlier experiments excluded the impactor, resulting in GSDs closer to the upper end of this range). The challenge aerosol is dried with a counterflow membrane dryer (PermaPure MD-700-48S-3) and passed through an impactor to remove large particles, then a bipolar [85]Kr neutralizer, before being mixed with dilution air and passing through a sample chamber containing a respirator or filtration media. (Some of the data presented for candidate N95 respirators was taken without the dryer.) A pair of scanning mobility particle sizers (SMPSs; composed of a subset of TSI model 3080, 3082, 3750, and 3752 differential mobility analyzers) and a pair of concentration particle counters (CPCs; composed of a subset of TSI models 3025A, 3082, 3776, and 3788), which were used to measure total number concentrations without size classification, are placed before and after the sample chamber to make measurements of the aerosol concentration upstream and downstream of the filters, respectively.

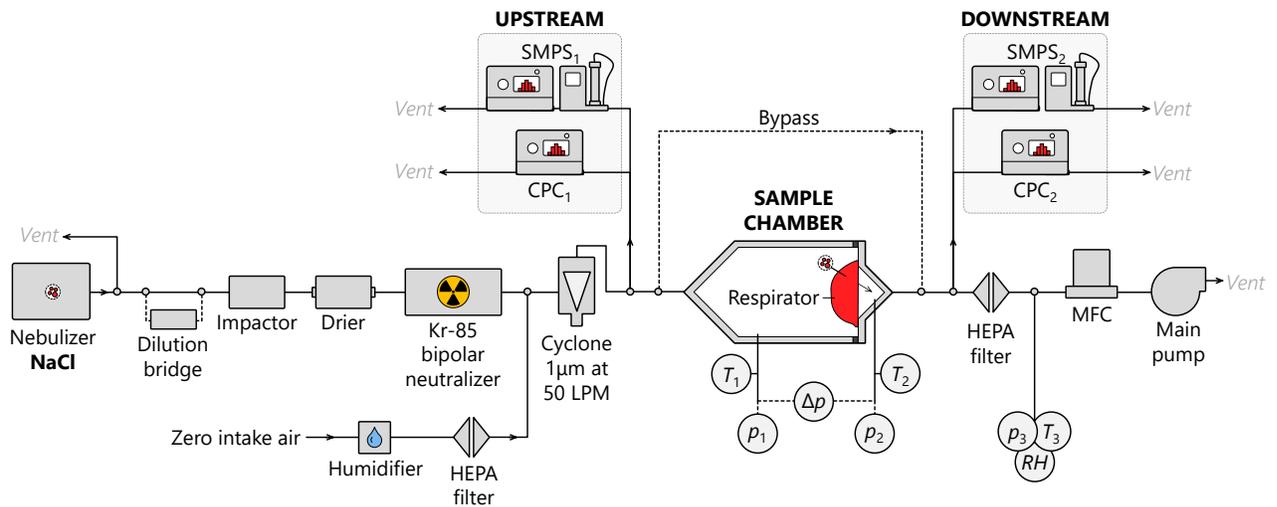

**Figure 3.** Schematic of the PFEMS system, composed of a pair of SMPSs and a pair of CPCs. The $p$, $T$, and $RH$ circles refer to pressure, temperature, and relative humidity sensors. Pressure sensors $p_1$ and $p_2$ were replaced by a differential pressure sensor prior, measuring $\Delta p$ directly, to measuring the candidate reference materials noted at the beginning of Sec. 5.

The PFEMS measurements are used to compute SPFE, NBFE, and MBFE. We report NBFE values from the PFEMS using Eq. (5), based on CPC data. NBFEs from integrated SMPS measurements (from a combination of Eqs. (4) and (5)) were consistent with the CPC results across the whole range of observed values (different by 1.2% of $P_n$ on average). We converted the PFEMS NBFEs to MBFEs using Eq. (8), with size distribution statistics ($d_{g,1}$, $d_{g,2}$, $\sigma_{g,1}$, and $\sigma_{g,2}$) computed from the SMPS data and $P_n$ computed from the CPC data.

Uncertainties for the PFEMS follow from linear error propagation (Corbin et al. 2021; Sipkens et al. Submitted). For SPFEs, this establishes that uncertainties rapidly expand in the tails of the challenge aerosol's particle size distribution. Overall, upstream number concentrations measured by the CPC are typically on the order of $5 \times 10^4$ counts, theoretically allowing measurements of NBFE up to 99.99% (before 2 standard deviations exceeds 100% of the penetration). For NBFE and MBFE, since the input number concentrations are relatively constant at this value, uncertainties in the penetration generally scale with the penetration itself and are thus uniform on a log-penetration scale. Typically, samples were tested for five minutes, averaging the measured quantities over this period, and resulting in total loading on the order of 0.01 to 0.3 mg, depending on the NBFE and the upstream aerosol properties (in particular the GSD and number concentration). Such loadings are well below that specified in the NIOSH N95 test method.

Filtration efficiencies are complemented by efforts to compute the most-penetrating particle size (MPPS) for each sample, computed from the measured SPFE curve. We estimate the MPPS using a probabilistic approach described in the Supplemental Information. Simulations employing a lognormal input challenge aerosol and lognormal penetration curve show that this method can predict the true MPPS within the stated uncertainties. Note that the MPPS was not always well-resolved. For example, in some instances the SPFE curve steadily declines over the entire observable range, such that the MPPS is at the edge of the distribution where uncertainties mask any ability to resolve the MPPS beyond the edge of the distribution. These points are denoted as such in subsequent figures (e.g., white-filled circles in Figure 4).

### 4.2. TSI 8130A

The TSI 8130A employs upstream and downstream photometers to determine PFE. The photometers measure light scattering from the particles, which is correlated to a mass. In effect, the instrument actually measures a scattering-based filtration efficiency, which is correlated with the MBFE. In practice, users are not presented with scattering information but with MBFEs reported by the instrument's embedded software based on the manufacturer's calibration. We report these values without modification, and compare them with the equivalent MBFE computed from the PFEMS. We report uncertainties in the TSI 8130A outputs from repeated measurements over time; standard filters, when available; or differences between two TSI 8130As, which we hereafter denote as TSI-1 and TSI-2 (see the Supplemental Information for a comparison). Data reported here correspond to an *initial* filtration efficiency, that is, the filtration after one minute of loading. Even these loadings are higher than that observed in the PFEMS system, which may result in differences between the measurements depending on the type of material (Barrett and Rousseau 1998). (While backwards extrapolation of the TSI 8130A data using points over the first five minutes is possible, this approach was ultimately not used for the data reported in this work.) For the mass concentrations employed in this study, this resulted in an average minimum loading of 1.5 mg of sodium chloride for a volumetric flow of 85 litres per minute (lpm). For higher MBFE, the initial loading will be larger, and vice versa.

Figure S1 in the Supplemental Information compares the MBFEs computed from the two TSI 8130A instruments. In the vast majority of cases, TSI-1 and TSI-2 agree within 25% of $P_m$. Note that, since these plots are logarithmic with respect to penetration, the results indicate that the MBFEs are in better absolute agreement for higher performing filtration media, consistent with expected trends (Sipkens et al. Submitted). Penetration measured by TSI-2 was 7% higher than TSI-1 on average, with a standard deviation in the difference between the penetrations of 15%.

# 5. Results

Measurements were made with the PFEMS and TSI 8130A systems across a range of materials. Data were roughly divided into three sets: (*i*) a range of N95 candidate respirators and filtration media, about which specific material properties are not generally known but that encompass more than 500 samples spanning 20%-99.8% NBFE; (*ii*) a subset of N95 candidate respirators from the previous category, for which dedicated measurements were made on both the PFEMS and a TSI 8130A for the same lot; and (*iii*) a set of candidate reference materials composed of melt-blown polypropylene that were controlled with respect to their basis weight (the mass per unit area of the material, typically expressed in $g/m^2$ or GSM) and degree of charging (by subjecting the material to electric fields of different voltages). The final category allows for some investigation of qualitative trends in PFE as a function of the underlying material parameters. Our measurements also include both as-received (unconditioned) and environmentally conditioned (85% ± 5% RH, 38 °C ± 2.5 °C for 25 h ± 1 h) samples. We refer the reader to Corbin et al. (2021) for discussion of a subset of this data within that context, as well as a discussion of differences between the NIOSH N95 and ASTM F2299/F2100 test methods.

## 5.1. PFEMS number-to-mass conversion

Figure 4 shows the relationship between NBFE and MBFE for experimental data across over 900 samples, with points coloured based on the observed MPPS. Note that the horizontal and vertical scales are logarithmic with respect to penetration but reversed such that filtration efficiency increases towards the upper, right corner. This scale was chosen for multiple reasons, including: that it better reflects trends in uncertainties, which are roughly proportional to the penetration for much of the domain (Sipkens et al. Submitted); that it better distributes the data across the plot area, which are clustered at high filtration efficiencies; and that it maintains an increasing filtration efficiency in moving towards the top-right. Four observations are evident from Figure 4.

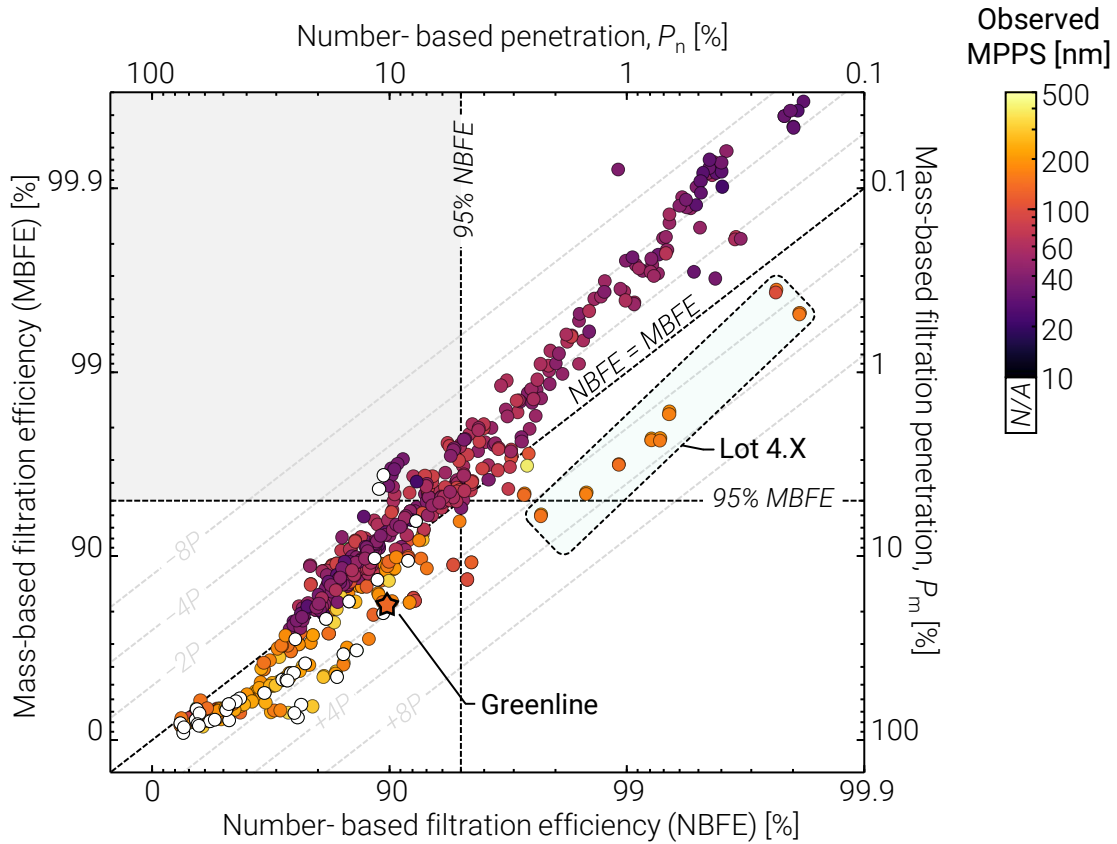

**Figure 4.** Mapping the relationship between NBFE and MBFE for PFEMS data taken on a wide range of filtration media and candidate N95 respirators, and coloured by the observed most-penetrating particle size (MPPS). White-filled circles indicate cases where estimates of the MPPS failed or were considered outside of a valid range. The challenge aerosol is sodium chloride with an approximate CMD of 75 nm and GSD between 1.65 and 1.86. Both the horizontal and vertical scales are logarithmic with respect to penetration. Faint, dashed lines correspond to multiples of the penetration (e.g., +8P, corresponds to when the penetration with respect to mass is eight times higher than the number-based quantity). Lot 4.X was a lot of particularly high pressure drop and basis weight filter media (dashed rectangle). Shaded region in the upper, left corner correspond to materials that have 95% MBFE (would pass by the corresponding test method), but not 95% NBFE.

First, there are many instances in which NBFE does not equal MBFE. This is indicative of significant variations in the SPFE, resulting from the fact that size-independent SPFE would necessarily yield NBFE = MBFE, as demonstrated schematically in Figure 2a. Note that NBFE = MBFE can also be achieved when the SPFE has a MPPS between the number and mass distributions, such that the opposite (that is the SPFE is necessarily flat when NBFE = MBFE) does not hold. This observation matches inspection of the SPFE curves and trends in single fiber efficiency models (Hinds 1999a).

Second, and related to the former, the MPPS has a significant impact on the difference between the NBFE and MBFE. Generally, MPPS > 80 nm (CMD of the challenge aerosol) results in a decrease when converting from NBFE to MBFE, while the converse is true for MPPS < 80 nm. This is supported by the theoretical discussion in Sec. 3.4. The difference between the NBFE and MBFE and the MPPS remains sufficiently scattered that a simple correlation is not possible.

Third, for the whole range of materials, there was a persistent relationship wherein high NBFE was associated with a smaller MPPS and thus yielded an increase in the filtration efficiency when converting to MBFE (upper-right sector of Figure 4). Conversely, poor NBFE materials often had a larger MPPS and thus saw a decline in the conversion from NBFE to MBFE (lower-left sector of Figure 4). In other words, the slope of a line through the data is steeper than the line of parity in Figure 4 and the relationship between MBFE and NBFE is inversely proportional to MPPS. The consistency of the trend across conditioned and unconditioned masks, a large range of NBFE, and a large number of lots is interesting. The materials in this study are typically non-wovens (such as melt-blown polypropylene), suggesting this trend may hold for this class of materials. Caution remains extrapolating such a trend more generally to all materials, particularly to materials for face coverings worn by the general public and across materials with various levels of electrostatic charge (Zangmeister et al. 2020; Rogak et al. 2021). However, most materials wherein MBFE > NBFE have an MPPS ~ 50 nm, which is consistent with previous observations (Eninger et al. 2008) and theoretical predictions of a shift downward in the MPPS (Huang et al. 2013; Wang and Otani 2013) for electret materials. As such, it is likely that the higher filtration materials often employ electrostatic capture to increase the PFE. Alternatively, the shift to lower MPPS could stem from thinner fibers with a finer mesh (Podgorski et al. 2006), though such differences may be more relevant in the sub-90% NBFE range.

Fourth, Figure 4 indicates the presence of a set of outliers in the 97%-99.8% NBFE level where MBFE < NBFE. These samples correspond to a single set of melt-blown polypropylene media (denoted here as Lot 4.X) with high PFEs, abnormally large MPPS, and exceptionally high pressure drops (breathing resistance of 200 – 500 Pa). Collectively, this suggests that Lot 4.X was a higher basis weight material, which relied more on mechanical filtration to achieve higher filtration efficiencies than the other materials in this NBFE range.

Overall, the results in Figure 4 demonstrate that there a significant number of samples from these respirators and filtration materials that exceed the 95% threshold for MBFE but have lower NBFE values, mostly in the 90-95% range. These are shown as a shaded region in the upper, left corner of Figure 4, and would pass the corresponding test method for MBFE, despite having a lower NBFE. This strengthens the position that standards should explicitly state whether quantities are to be mass- or number-based, as there are sufficient instances where a mask may pass by mass but not by number, particularly for electret materials. We also note that the converse, where NBFE > 95% but the MBFE was not, was very rare for the range of samples considered in this work. While this region may be relevant if the high performing candidate respirators are sanitized for reuse and lose their electret properties (though the performance of respirators under these conditions is an active area of research, e.g., (Carrillo et al. 2020; Czubryt et al. 2020; Everts et al. 2021; Jatta et al. 2021)), this may reduce the filtration properties sufficiently that both the NBFE and MBFE will be below 95%. Further, samples in this quadrant are likely to be uncharged, instead relying on mechanical filtration to achieve high filtration. As a result, these materials are likely to be associated with higher pressure drops that make these materials poor candidates for respirators and likely to fail pressure resistance requirements in many standards. As such, NBFE would remain a reasonable surrogate for gauging compliance under most circumstances.

Figure 5 further examines the impact of the properties of the filtration media by considering the candidate reference material subset of the data, for which the charge state and basis weight of the material are controlled. Here, the uncharged materials correspond (*i*) to the lower NBFE cases and (*ii*) to the cases where the MPPS is large, such that MBFE is less than NBFE. Examination of size-resolved filtration curves (cf. Figure 5*b*) affirms this, with the uncharged materials exhibiting a distinctly different SPFE profile relative to the charged materials. Within each cluster in Figure 5*a* (or the shaded regions in Figure 5*b*), higher basis weights yield higher quality filtration media within the given charge state data cloud (higher in the shaded regions), consistent with an increase in solidity, fiber size, and/or media thickness. For the uncharged samples, this results in an increasing gap between the NBFE and MBFE. Greenline media (a commercial sample used by the TSI 8130A instrument as an internal reference) are uncharged and have a much higher MPPS, resulting in a decrease in the conversion from NBFE to MBFE. The resultant position in Figure 5*a* is consistent with an uncharged media with higher mechanical filtration than most of the candidate reference materials considered in this study. This affirms statements made in connection with Figure 4, where the higher performing materials were hypothesized to be electret materials.

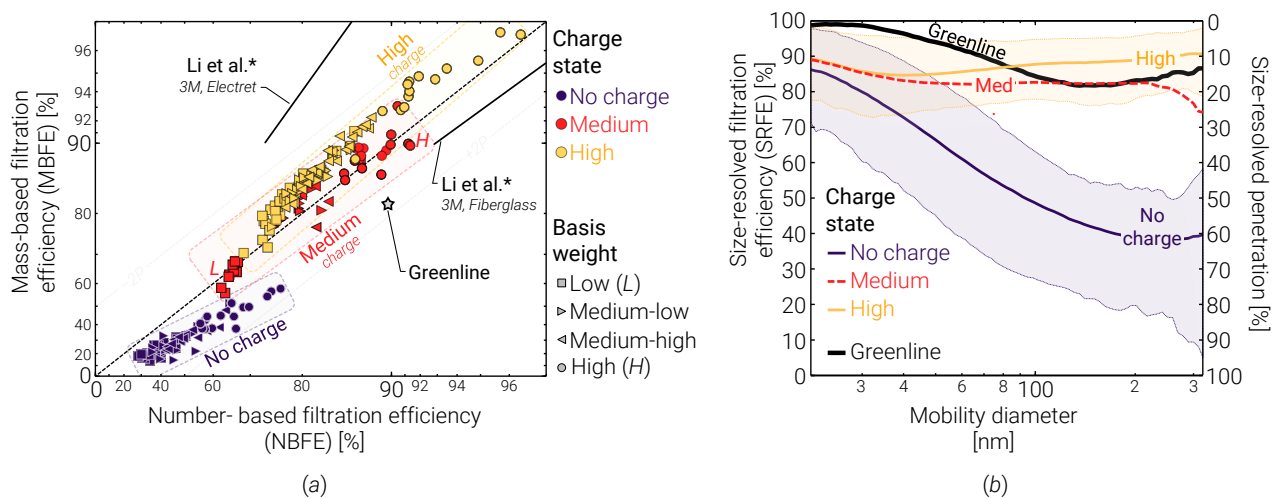

**Figure 5.** Examination of the SPFE, NBFE, and MBFE for a range of candidate reference filtration media where material the charge state and basis weight have been controlled. (*a*) NBFE versus MBFE for PFEMS data across a series of candidate reference materials, coloured by charge state and with shapes corresponding to different basis weights. Shaded regions group common charge states, while "H" and "L" mark the high and low basis weight points for the medium charge state. (*b*) SPFEs from the PFEMS for a range of candidate reference materials plus single-layer Greenline media, stated as the mean and 5 to 95% percentiles (shaded regions) of the materials for each charge state. Percentiles are excluded for the medium charge state in (*b*) for clarity. The observed decrease at the upper diameter end of the medium charge state case in (*b*) corresponds to a region of expanding uncertainties for those curves and is non-physical. Charge state refers to the voltage applied to the material during manufacture.

Such results are also largely consistent with Li et al. (2012), where MBFE was measured using a TSI 8130 and NBFE using a pair of CPCs. In that study, researchers had varied input distributions across the range of NBFEs, varying the CMD from 36 to 68 nm and the GSD from 1.5 to 1.93 to realize a broader range of PFEs. This limits comparison to the present study, where the aerosol size distribution was more consistent and a broader range of materials was considered. Nevertheless,

Li et al. (2012) considered two types of materials: electret and uncharged, fiberglass (akin in terms of material class to Greenline media) materials, which had MPPSs of ~50 nm and 160 nm, respectively. In all cases, they observed that fiberglass samples had MBFE < NBFE, consistent with the MPPS being much larger than the CMD of the challenge aerosol. Conversely, their electret material saw MBFE ≫ NBFE, consistent with the MPPS being smaller than the CMD. The trend observed in the uncharged fiberglass samples measured by Li et al. (2012) is also consistent with the outlier Lot 4.X in Figure 4. These trends are also consistent with earlier studies that did not discuss NBFE or MBFE but showed that electrostatic capture is least efficient for smaller particles (≈ 100 nm aerodynamic diameter) under the conditions relevant to respirator testing (Corbin et al. 2021).

### 5.2. Experimental validation of PFEMS and TSI 8130A MBFEs

Figure 6 indicates the relationship between the PFEMS MBFE computed using the aforementioned method and the TSI 8130A MBFEs. Note that, as before, both axes are on a logarithmic scale with respect to penetration, to space out the high filtration cases. On this scale, relative uncertainties in the penetration are roughly constant across the entire domain (which is equivalent to lower absolute uncertainties in the high MBFE points, i.e., absolute uncertainties are lower at 99% than they are at 30% MBFE).

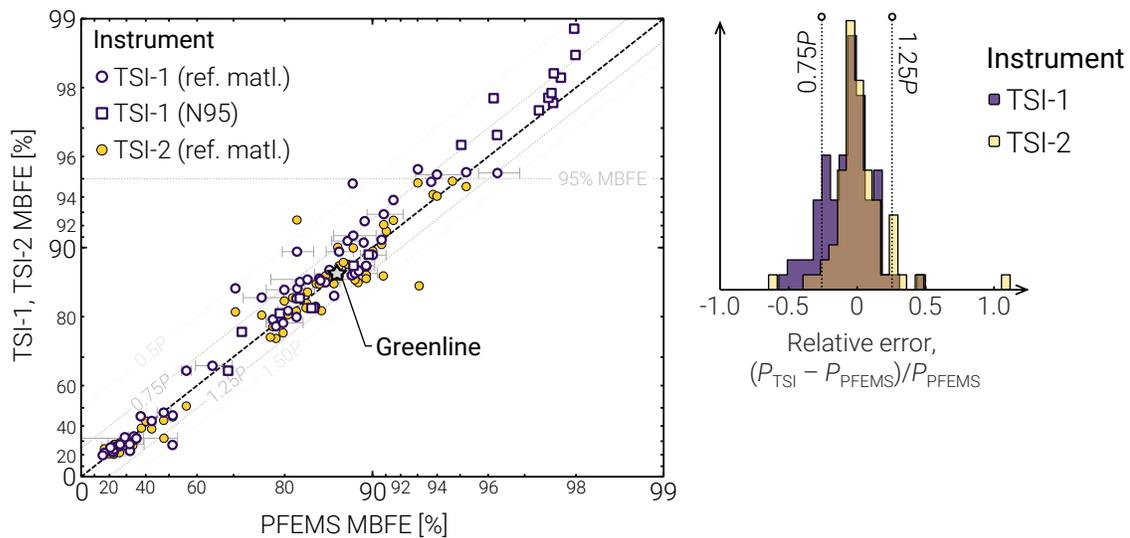

**Figure 6.** Parity plot and histogram showing the relationship between the MBFE computed using the PFEMS number-based systems versus the MBFE output of two TSI 8130As (i.e., TSI-1 and TSI-2) for the candidate reference materials (denoted ref. matl.). Error bounds correspond to the standard deviation across multiple (typically three) repeats for the PFEMS and are excluded from the comparison to TSI-2 for clarity. The symbols for TSI-1 refer to whether the data corresponds to (*circles*) a candidate reference material (cf. Figure 5) or (*squares*) a set of candidate N95 respirators. Inset shows a histogram of the relative errors in the TSI 8130A measurements relative to the PFEMS.

Overall, the MBFE is highly correlated between all three instruments, indicating consistency between these two types of instruments and the conversion of the PFEMS data to MBFE removes structure in the error between the two classes of systems. The histogram of the ratio of the TSI 8130A MBFE to that derived from the PFEMS generally affirms this observation, but also shows some difference between the two TSI 8130As, which typically results in outliers in the tails of the different histograms. As a result, the PFEMS MBFEs often exist between the MBFEs output by the TSI 8130As, suggesting that the PFEMS may be more consistent and that much of the scatter here results from variability between the TSI 8130A instruments. This may stem from insensitivities of the photometer output to the small particles present in the challenge aerosol (Eninger et al. 2008), limiting the accuracy of the calibration procedures, particularly when the size distribution is not lognormal, or to small shifts in the nebulizer output from the TSI 8130As. A majority of samples yielded similar results between all three devices, yielding inter-device errors on the order of 25% (one standard deviation) of the mass-based penetration.

Greenline media MBFEs, in particular, showed excellent agreement between the PFEMS and the TSI 8130As, which may support the above hypothesis that inaccuracy due to small particles plays a role in the scatter of Figure 6 and may support improvements to the calibration procedures associated with these measurements.

## 6. Conclusions

This work presents data from over 900 samples examining the relationship between the NBFE and MBFE. We have demonstrated consistencies between the output of a number-based system (PFEMS) and the mass-based filtration efficiencies (MBFEs) output by the TSI 8130A, the instrument most often used in conjunction with the NIOSH TEB-APR-STP-0059 standard.

Empirical observations suggest a common scenario in which high NBFE materials are often associated with (*i*) electret materials, as expected; (*ii*) a smaller MPPS; and (*iii*) an increase when converting from NBFE to MBFE (a consequence of the smaller MPPS). Data in which MBFE > 95% but NBFE < 95%, were sufficiently common to indicate that documentary standards should explicitly state whether filtration efficiencies are mass-based. Cases where NBFE > 95% but MBFE < 95% were infrequent, suggesting that the NBFE may be used to assess performance for mass-based standards under many conditions. While uncharged materials may disrupt this simple interpretation, these materials are more likely to fail pressure drop requirements, due to the reliance on mechanical filtration.

Absolute differences between the output of the number-based system and that of the TSI 8130A decrease as the PFE increases. Relative errors in the penetration were approximately constant across a wide range of PFEs. Larger variations appear to be present between two TSI 8130A systems than between the TSI 8130As and the number-based system. Generally, consistencies between all three instruments were within 25% (one standard deviation) of the measured mass-based penetration.

We also demonstrate the equivalence of SMPS-based and calibrated photometer systems, when both are expressed as MBFEs. We emphasize the different particle types and how size-resolved filtration efficiencies must be used to convert between the quantities, as they are not generally equivalent.

# Funding

This work has been funded by the Public Health Agency of Canada (PHAC) and by Pillar 4 of the National Research Council Canada (NRC) Pandemic Response Challenge Program.

# Acknowledgements

We thank the entire NRC COVID-19 Respirator Testing Team for their contribution to the measurements discussed here, including: Adam Willes, Adrian Simon, Aiden Korycki-Striegler, Ali Ghaemi, Allison Sibley, Amor Duric, Amr Said, Anabelle Bourgeois, Andason Cen, Andre Cantin, Andrew Oldershaw, Apoorv Shah, Brett Smith, Bryan Muir, Cam Lebrun, Cindy Jiang, Chantal Prévost, Christina Brophy, Dan Clavel, Dave Angelo, David Kennedy, Deval Patel, Doug Mackenzie, Douglas McIntyre, Erhan Dikel, Flamur Canaj, Gang Nong, Garnet McRae, Getho Eliodor, Greg Nilsson, Harold Parks, Isabelle Rajotte, James Renaud, James Saragosa, Jason Brown, Jean Dessureault, Jeff Tomkins, Jeremy Macra, Joshua Marleau-Gillette, Kari McGuire, Kimberly Moore, Krystal Davis, Laura Forero, Lyne St-Cyre, Marilyn Azichoba, Mario Carrière, Mark Vuotari, Michel Levesque, Michael Ryan, Mladen Jankovic, Nicholas Wise, Nikolas Angelo, Ovi Mihai, Peter Hanes, Richard Agbeve, Samuel Camiré, Simon-Alexandre Lussier, Sonia Mutombo, Stacey Lee, Stephane Lapointe, Stephanie Gagné, Steve Kruithof, Thierry Lavoie, Tomi Schebywolok, Tyler Hunter, Vicki Wang, Xigeng Zhao, Yanen Guo.

# ORCID

Timothy A. Sipkens · https://orcid.org/0000-0003-1719-7105
Gregory J. Smallwood · https://orcid.org/0000-0002-6602-1926
Joel C. Corbin · https://orcid.org/0000-0002-2584-9137
Prem Lobo · https://orcid.org/0000-0003-0626-6646
Jalal Norooz Oliaee · https://orcid.org/0000-0003-3839-2313